\documentclass[a4paper,12pt]{article}
\usepackage{amsmath,amssymb}
\usepackage{hyperref}
\usepackage{graphicx}
\usepackage{mathrsfs}
\usepackage{float}
\usepackage{ulem}

\newcommand{\be}{\begin{equation}}
\newcommand{\ee}{\end{equation}}
\newcommand{\f}{\frac}

\newcommand{\bea}{\begin{eqnarray}}
\newcommand{\eea}{\end{eqnarray}}
\newcommand{\ba}{\begin{align}}
\newcommand{\ea}{\end{align}}

\newcommand{\ra}{\rangle}
\newcommand{\la}{\langle}

\begin{document}

\begin{titlepage}

\vspace{.4cm}
\begin{center}
\noindent{\Large \textbf{Mutual information of excited states and relative entropy of two disjoint subsystems in CFT}}\\
\vspace{1cm}
 Tomonori Ugajin$^{b,}$\footnote{ugajin@kitp.ucsb.edu}

\vspace{.5cm}
\vspace{.5cm}
  {\it
 $^{b}$Kavli Institute for Theoretical Physics, University of California, \\
Santa Barbara, 
CA 93106, USA\\
\vspace{0.2cm}
 }
\end{center}


\begin{abstract}
In this paper, we first  study mutual information of excited states in the small subsystem size  limit in generic conformal field theory. We then  discuss relative  entropy of two disjoint subsystems  in the same limit.
\end{abstract}

\end{titlepage}

\tableofcontents
\section{Introduction}

Quantum information theoretic approaches have been uncovering  several profound properties of quantum field theory.  
 In these approaches, nice features of  mutual information and relative entropy, such as positivity  or monotonicity play a crucial role in deriving these results \cite{Casini:2004bw, Casini:2012ei,Casini:2015woa, Casini:2016fgb, Casini:2016udt, Faulkner:2016mzt,
Wall:2010cj,Wall:2011hj,Bousso:2014sda,Bousso:2014uxa}.  Furthermore, these properties 
 can be used to constrain the dual gravitational dynamics and spacetime structure through holography.  
For example,  positivity of  relative  entropy implies  bulk linearized equations of motion \cite{Lashkari:2013koa,Faulkner:2013ica,Nozaki:2013vta} and constraints  bulk matter theory 
\cite{Lin:2014hva, Lashkari:2014kda, Lashkari:2015hha,Lashkari:2016idm, Lin:2016fua}.

\vspace{0.1cm}

Mutual information $I(A,B)$ in quantum field theory measures the correlations between two regions A and B.   More precisely, this quantity gives an upper bound on correlations between A and B \cite{boundMI},
\be
I(A,B) \geq \frac{(\langle O_{A} O_{B} \rangle -\langle O_{A} \rangle  \langle O_{B} \rangle)^2}{||O_{A}||^2 ||O_{B}||^2},
\ee
where $O_{A}$ and $O_{B}$ denote bounded operators on the region A and B respectively.  
In \cite{Calabrese:2009ez,Calabrese:2010he,Headrick:2010zt}  the mutual information of vacuum in two dimensional conformal field theory was computed in the limit where the distance between 
A and B is large. \footnote{This is equivalent to the small subsystem size limit $|A|, |B| \rightarrow 0$ with the distance between  them kept fixed.} This was done by  expanding  the four point function of the twist operators using their OPE, and gathering the leading order contributions in this limit.  
This calculation is generalized to CFT in higher dimensions \cite{Cardy:2013nua,Agon:2015ftl},  by introducing the notion of a higher dimensional twist operator \cite{Cardy:2013nua}.  A nice feature of this  formalism is that it makes the relation between 
2d calculation and it's higher dimensional counterpart  transparent.

In the first part of this paper, we calculate the mutual information of an arbitrary excited state $| V \rangle$  in generic conformal field theory in the small subsystem size limit. We derive this by computing a correlation  function involving twist operators 
and operators representing the excited state. The result is given by , 

\be
I(A,B) = (l_{A})^{2\Delta} ( l_{B})^{2\Delta} \frac{\Gamma(\f{3}{2})\Gamma(2\Delta+1)}{2\Gamma(2\Delta+\f{3}{2})}  \Biggl[ \langle V|  \mathcal{O}_{\alpha} \mathcal{O}_{\beta} |V \rangle- \langle V| \mathcal{O}_{\alpha} |V \rangle \langle V| \mathcal{O}_{\beta} |V \rangle \Biggr]^2 \label{eq:main},
\ee  
where $\mathcal{O}$ is the lightest non vacuum operator in the CFT\footnote{In this paper, we assume that  this operator $\mathcal{O}$ is a scalar. It should be straightforward to generalize this to other cases where  $\mathcal{O}$ has spin. }, and $\Delta$ is the conformal dimension of it. $(l_{A}, \alpha)$ and $(l_{B}, \beta)$ denote  (radius, location) of the subsystem A and B respectively.  Note that we can uniquely specify the locations $\alpha, \beta$ since we are considering the small subsystem size limit.
The result is essentially given by  square of the connected two point function of the operator $\mathcal{O}$ evaluated on the excited state $| V \rangle$ .

\vspace{0.1cm}

Relative entropy $S( \rho|| \sigma)$ measures the distance between two density matrices $\rho$ and $\sigma$, and is defined by

\be
S( \rho|| \sigma)= {\rm tr} \rho \log \rho -  {\rm tr} \rho \log \sigma.
\ee
When the subsystem A is a single connected region, the relative entropy $ S_{A}(\rho_{V}|| \rho_{W})$ between two reduced density matrices $\rho_{V}, \rho_{W}$ of excited states $| V \rangle$, $ |W \rangle$,  $\rho_{V} ={\rm tr }_{A^{c}}|V \ra \la V| $
 $\rho_{V} ={\rm tr }_{A^{c}}|W \ra \la W| $ was computed  in the small subsystem limit, both  in 2d CFT\cite{Sarosi:2016oks} and CFT in higher dimensions \cite{relhigerpep}   by using the replica trick introduced in \cite{Lashkari:2015dia} .

\vspace{0.1cm}

In the second part of this paper, we derive a general formula for  the relative entropy $S_{A \cup B}(\rho_{V}|| \rho_{W})$ in the small interval limit $l_{A}, l_{B} \rightarrow 0$.  We do this by reading off the form of 
the modular Hamiltonian of the excited state $K^{W}_{A \cup B}$ from the mutual information result of the first part. 

\vspace{0.1cm}

The organization of this paper is as follows: In section \ref{section: 2dmi}, and  section \ref{section:smie} , we discuss the mutual information of an excited state in the small subsystem size limit in 2d CFT.  In section \ref{section:hdmi} we generalize this 
result to higher dimensions. In section \ref{section: hdrel}, we derive an expression of the relative entropy $S( \rho_{V}|| \rho_{W})$ of the disjoint subsystems in the same limit by using the results of the previous 
two sections.

\section{ Mutual information in 2d CFT} \label{section: 2dmi}
In this section, we focus the on the  mutual information,
\be
I(A:B)=S_{A}+S_{B}-S_{A\cup B}, \label{eq:mi}
\ee
of an excited state in 2d CFT. We consider a CFT defined on a cylinder $S^{1} \times \mathbb{R}$, and take the subsystems $A=[ l_{1}, l_{2}]$, $B=[l_{3},l_{4}]$ on a time slice $S^{1}$ of the cylinder.   

We compute the mutual information (\ref{eq:mi}) by using the replica trick,
\be
I(A:B)= \lim_{n \rightarrow 1}\f{1}{n-1} I_{n} (A:B), \label{eq:repin}
\ee
with
\begin{align}
I_{n}(A:B)=\log {\rm tr} \;\rho^{n}_{A \cup B} -\log {\rm tr}\; \rho^{n}_{A}-\log {\rm tr} \;\rho^{n}_{B}. \label{eq:replicami}
\end{align}

Each term in (\ref{eq:replicami}) can be expressed by a path integral on an n-sheeted cylinder with the cut on the subsystem. 
By the exponential map $w=e^{iz}$, each  cylinder sheet  is mapped to the plane, and employing the orbifold description of the theory on the n sheeted plane $\Sigma_{n}$, one can express them as  the correlation functions involving the twist operators $\sigma_{n}, \tilde{\sigma}_{-n}$ \cite{Calabrese:2004eu},

\begin{align}
 {\rm tr} \;\rho^{n}_{A \cup B}&= \langle V_{n}(\infty) \sigma_{n} (z_{1}) \tilde{\sigma}_{-n}(z_{2}) \sigma_{n} (z_{3}) \tilde{\sigma}_{-n}(z_{4})V_{n}(0) \rangle, \quad V_{n} \equiv V^{\otimes n}, \\[+10pt] 
 {\rm tr} \;\rho^{n}_{A}&= \langle V_{n}(\infty) \sigma_{n} (z_{1}) \tilde{\sigma}_{-n}(z_{2}) V_{n}(0) \rangle, \quad  {\rm tr} \;\rho^{n}_{B}= \;\langle V_{n}(\infty) \sigma_{n} (z_{3}) \tilde{\sigma}_{-n}(z_{4}) V_{n}(0) \rangle,
\end{align}
where 
\be 
z_{1}=e^{il_{1}}, z_{2}=e^{il_{2}}, \quad z_{3}=e^{il_{3}}, \quad z_{4}=e^{il_{4}}.
\ee

By using these, we obtain, 
\be
 I_{n}(A:B) = \log \left[\f{\langle V_{n}(\infty) \sigma_{n} (z_{1}) \tilde{\sigma}_{-n}(z_{2}) \sigma_{n} (z_{3}) \tilde{\sigma}_{-n}(z_{4})V_{n}(0) \rangle }{\langle V_{n}(\infty) \sigma_{n} (z_{1}) \tilde{\sigma}_{-n}(z_{2}) V_{n}(0) \rangle \langle V_{n}(\infty) \sigma_{n} (z_{3}) \tilde{\sigma}_{-n}(z_{4}) V_{n}(0) \rangle}\right]. \label{eq:minont}
\ee

\section{Small interval expansion}
\label{section:smie}
\subsection{Substituting the OPEs}
The objective of this section  is to find the leading behavior of the mutual information (\ref{eq:mi}) in the small 
subsystem limit,  $l_{12} \equiv l_{1} - l_{2} \rightarrow 0, \quad  l_{34}\equiv l_{3} - l_{4} \rightarrow 0 $ with $l_{3}$ kept fixed.
To see the leading behavior, we use the OPE of the twist operators 
\cite{Calabrese:2009ez,Calabrese:2010he,Headrick:2010zt}, 
\be 
\sigma_{n}(z_{1}) \tilde{\sigma}_{-n}(z_{2}) = \langle \sigma_{n}(z_{1}) \tilde{\sigma}_{-n}(z_{2}) \rangle \left\{ 1+  \sum^{n/2}_{j=1}  C_{\sigma_{n}\tilde{\sigma}_{-n}}^{T_{j}} T^{j}(z_{2}) (l_{12})^{\Delta_{T}} + \cdots \right\},
\ee
with
\begin{align} 
T^{j}(z_{2}) &= \left[ \mathcal{O}(z_{2}) \otimes I^{\otimes (j-1)}\mathcal{O}(z_{2}) \otimes I^{\otimes (n-j+1)} \right]_{sym} \nonumber \\[+10pt]
&\equiv \f{1}{n} \sum_{p_{1}} B_{z_{2}} (p_{1},p_{1}+j),
\end{align}
where $\mathcal{O}$ is the lightest non vacuum state of the seed theory with the conformal dimension $\Delta$,  therefore  $\Delta_{T} =2 \Delta$, and we have defined
\be
B_{z_{2}} (p_{1},p_{1}+j) \equiv I^{\otimes (p_{1}-1)}\otimes \mathcal{O} (z_{2})\otimes I^{\otimes (j-1)} \otimes  \mathcal{O}(z_{2}) \otimes I^{\otimes (n-p_{1}-j)}.
\ee
 There is a similar OPE for $\sigma_{n}(z_{3}) \tilde{\sigma}_{-n}(z_{4})$.

\vspace{0.2cm}

Substituting these OPEs into (\ref{eq:minont}), we get,  
\be
I_{n}(A:B)= (l_{12})^{2\Delta} (l_{34})^{2\Delta}\sum_{a,b}^{\f{n}{2}}  C_{\sigma_{n}\tilde{\sigma}_{-n}}^{T_{a}}  C_{\sigma_{n}\tilde{\sigma}_{-n}}^{T_{b}}  \langle V_{n}(\infty) T^{a}(z_{2}) T^{b}(z_{4}) V_{n}(0) \rangle, 
\label{eq: fourpt}
\ee
in the small subsystem size limit, $l_{12}, l_{34} \rightarrow 0$.
Notice that since 
\be
\sum^{n}_{j} C^{T_{j}}_{\sigma_{n}\tilde{\sigma}_{-n}} T_{j} = \sum_{(m, k) =1,m\neq k}^{n} C^{B(m,k)}_{\sigma_{n}\tilde{\sigma}_{-n}}\; B(m,k),
\ee

it can be written in terms of $B(j,k)$,

\be
I_{n}(A:B)=\f{1}{4} (l_{12}l_{34})^{2\Delta} \sum_{(j,p_{1}) =1, j \neq p_{1} }^{n} \sum_{(k,p_{2}) =1, k \neq p_{2} }^{n} C^{B(j , p_{1})}_{\sigma_{n}\tilde{\sigma}_{-n}} C^{B(k, p_{2})}_{\sigma_{n}\tilde{\sigma}_{-n}} 
\; \langle V_{n}(\infty) B_{z_{2}}(p_{1}, j) B_{z_{4}} (p_{2}, k) V_{n}(0) \rangle. \label{eq:mutn}
\ee

\subsection{ Computation of $ I_{n}(A:B)$}
The strategy to compute (\ref{eq:mutn}) is, first computing  the sum  with respect to $j$ and $k$
\be
I_{p_{1} p_{2}}=\sum_{k=1, k \neq p_{1}}^{n}\sum_{j=1, j \neq p_{2}}^{n} \; \langle V_{n}(\infty) B_{z_{2}}(p_{1}, j) B_{z_{4}} (p_{2}, k) V_{n}(0) \rangle \; C_{\sigma_{n}\tilde{\sigma}_{-n}}^{B(j,p_{1})} C_{\sigma_{n}\tilde{\sigma}_{-n}}^{B(k,p_{2})} \equiv \sum_{k=1, k \neq p_{1}}^{n}\sum_{j=1, j \neq p_{2}}^{n}  I^{k,j}_{p_{1} p_{2}},
\ee
while keeping $p_{1}, p_{2}$ fixed, then perform the sum with respect to $p_{1},p_{2}$. The precise form of the summand $I^{k,j}_{p_{1},p_{2}}$ depends on the value of the  indices, for example $k=j_{2}$ or $k \neq j_{2}$.  We classify them and perform the sum carefully in Appendix A.

\vspace{0.2cm}

 After these calculations,  we get following  expression of  $I_{n}(A:B)$,
\begin{align}
 I_{n}(A:B)&= \f{1}{4} (l_{12})^{2\Delta} (l_{34})^{2\Delta} \left(\sum _{p_{1} \neq p_{2}} I_{p_{1}, p_{2}} + \sum_{p_{1}} I_{p_{1}, p_{1}} \right) \nonumber \\[+10pt]
&= \f{1}{2}  (l_{12})^{2\Delta} (l_{34})^{2\Delta} \left[  \langle  \mathcal{O}_{\alpha} \mathcal{O}_{\beta} \rangle- \langle \mathcal{O}_{\alpha} \rangle \langle \mathcal{O}_{\beta} \rangle \right]^2  \sum_{k=1, k\neq p_{1}}^{n} C(p_{1}-k)^2, \label{eq:inex}
\end{align}

here we introduced  the simplified notations,
\be
C(p_{1}-p_{2}) \equiv C^{B(p_{1},p_{2})}_{\sigma_{n} \tilde{\sigma}_{-n}}\quad   \langle \mathcal{O}_{\alpha} \mathcal{O}_{\beta} \rangle \equiv \langle V(\infty)  \mathcal{O} (z_{2})\mathcal{O} (z_{4}) V(0) \rangle, \quad \la \mathcal{O}_{\alpha} \ra \equiv  \la V(\infty)  \mathcal{O}_{\alpha} V(0) \ra.
\ee
and we have omitted terms  which are not surviving the final  $n \rightarrow  1$ limit in (\ref{eq:repin}).

The $n \rightarrow 1$ limit of  (\ref{eq:inex}) can be easily taken by using the formula 
\cite{Calabrese:2009ez,Calabrese:2010he}, 
\be
f(\Delta,n)= \sum_{k=1, k\neq p_{1}}^{n} C(p_{1}-k)^2 =\sum^{n-1}_{k=1} \f{1}{\left(2n\sin \f{\pi k}{n}\right)^{4\Delta}} \rightarrow (n-1) \f{\Gamma(3/2)\Gamma(2\Delta+1)}{2^{4\Delta}\Gamma(2\Delta+3/2)}, \label{eq:anacon}
\ee
therefore we get, 
\be
I(A,B) = (l_{A})^{2\Delta} ( l_{B})^{2\Delta} \frac{\Gamma(\f{3}{2})\Gamma(2\Delta+1)}{2\Gamma(2\Delta+\f{3}{2})}  \Biggl[ \langle  \mathcal{O}_{\alpha} \mathcal{O}_{\beta} \rangle- \langle \mathcal{O}_{\alpha} \rangle \langle \mathcal{O}_{\beta} \rangle \Biggr]^2 \label{eq:finalresult}.
\ee

Here we introduced  radii  $l_{A}, l_{B}$ of the subsystems  $2l_{A} \equiv l_{12}, 2l_{B}  \equiv l_{34}$.

\section{ Mutual information in  higher dimensions} 
\label{section:hdmi} 

In this section we discuss the  higher dimensional generalization of the above 2d calculation, by using   higher dimensional  twist operators $\Sigma_{A}$ developed recently in  \cite{Cardy:2013nua,Agon:2015ftl,Hung:2014npa}. 

We will see that  the result (\ref{eq:finalresult}) remains to be true  in this case, once we interpret $l_{A},l_{B}$ as radii of the regions $A,B$.

\subsection{Setups}

We start from a d-dimensional conformal field theory on a cylinder $S^{d-1} \times \mathbb{R}$. We define the standard metric on this manifold, 
\be
ds^2= dt^{2} +(d\theta^2 + \sin^2 \theta d\Omega_{d-2}^2 ),
\ee

where $d \Omega_{d-2}^2$ denotes the metric on the d-2 dimensional sphere $S^{d-2}$. 

We choose the subsystem to be the union of two disjoint cap like regions $A \cup B$ on the spatial manifold $S^{d-1}$ at $t=0$.  We again assume that two regions $A$ and $B$ are small, therefore  we can define radii $l_{A}, l_{B}$ of the regions.  

 As in two dimensions, the trace of the reduced density matrix  ${\rm tr} \; \rho^{n}_{A \cup B}$
of an excited state $|V \rangle$  is given by the path integral on the n-sheet cover  of $S^{d-1} \times \mathbb{R}$ with the cut on the subsystem $A\cup B$.

It is again useful to map the replica manifold to the  n-sheet plane $C^{(n)}_{A \cup B} (\mathbb{R}^{d}) $ with the radial coordinate $r$, by the conformal map $t=\log r$. On the n-sheet plane, two  regions A and B are mapped to  regions in $r=1$ sphere, the excited states are located at $r=0, \infty$.  By using  state operator correspondence, we obtain,
\be
{\rm tr} \; \rho^{n}_{A \cup B}= Z^{(n)}_{A\cup B} \; \langle \prod^{n}_{j=1} V(\infty_{j}) V(0_{j}) \rangle_{C^{(n)}_{A \cup B} (\mathbb{R}^{d})}, \label{eq:highdimtra}
\ee
where $ Z^{(n)}_{A\cup B}$ denotes  the partition function on the n-sheet plane  $C^{(n)}_{A \cup B}  (\mathbb{R}^{d}) $. 

\subsection{ Higher dimensional twist operators} 

In this section we briefly review the concept of a higher dimensional twist operator \cite{Cardy:2013nua}. 
 A higher dimensional twist operator $\Sigma_{A}$ of  a region A is a non local operator defined in the n copies of the original CFT ( below this theory is denoted by $CFT^{\otimes n}$), and satisfies 
\be
\langle X \rangle_{C^{(n)}_{A}} =  \f{(Z^{(1)}_{A})^{n}}{Z^{(n)}_{A}}\langle X  \Sigma_{A}\rangle_{CFT^{\otimes n}},
\ee
where  $X$ is a product of local operators of the CFT on $C^{(n)}_{A}$. Note that we can  naturally  interpret $X$ to be an operator of $CFT^{\otimes n}$.

In the small subsystem size limit $|A| \rightarrow 0$, one can expand the twist operator by a set of local operators of  $CFT^{\otimes n}$  on the region A \cite{Cardy:2013nua} ,
\be
\Sigma_{A} =\f{Z^{(n)}_{A}}{(Z^{(1)}_{A})^{n}} \sum_{\{k_{j}\}} C^{A}_{\{k_{j}\}} \otimes^{n}_{j=1} \mathcal{O}_{k_{j}}(r_{A}), \label{eq:twistop}
\ee

Furthermore, the coefficient $C^{A}_{\{k_{j}\}}$ is given by the correlation function on $C^{(n)}_{A}$, 
\be
C^{A}_{\{k_{j}\}} =\lim_{r \rightarrow \infty} \langle \prod^{n}_{j=1} \mathcal{O}_{k_{j}} (r) \rangle_{C^{(n)}_{A}}.
\ee

When the original theory is defined on a conformally flat space, these coefficients are related to  correlation functions on $S^{1} \times H^{d-1}$, where $H^{d-1}$ is $d-1$ dimensional hyperbolic space \cite{Agon:2015ftl}.  If we pick up the first few terms in the small subsystem size expansion (\ref{eq:twistop}), we get

\be
\Sigma_{A} =\f{Z^{(n)}_{A}}{(Z^{(1)}_{A})^{n}} \left( 1+\f{1}{2} \sum^{n}_{(p_{1},j)=1,j\neq p_{1}} C^{A}_{(p_{1},j)}B_{r_{A}}(p_{1}, j)  +\cdots \right), \label{eq:exphightw}
\ee
where is $B_{r_{A}}(p_{1},j) $ is defined by 
\be
B_{r_{A}}(p_{1},j) = I^{\otimes (p_{1}-1)} \otimes \mathcal{O}(r_{A})\otimes I^{\otimes (j-p_{1}-1)} \otimes \mathcal{O}(r_{A}) \otimes I^{\otimes n-j},
\ee
and $\mathcal{O}$ is the lightest non vacuum operator with the conformal dimension $\Delta$.
From the conformal map of interest, we can specify the subsystem size dependence of the coefficient \cite{Agon:2015ftl},  
\be
C^{A}_{p_{1},j} =(2l_{A})^{\Delta} \tilde{C}_{p_{1},j},
\ee
and $\tilde{C}_{p_{1},j}$ is independent of $l_{A}$. 

\subsection{ The expression of $I_{n}(A,B)$}
By using the twist operators $\Sigma_{A}, \Sigma_{B}$, we can write $ {\rm tr}\; \rho_{V}^{n} (A \cup B)$ in term of the correlation function in $CFT^{\otimes n}$

\be
{\rm tr}\; \rho_{V}^{n} (A \cup B)= \langle V^{\otimes n}(\infty) \Sigma_{A} \Sigma_{B} V^{\otimes n} (0) \rangle_{CFT^{\otimes n}}.
\ee

There are similar expressions of  ${\rm tr} \rho_{V}^{n} (A), {\rm tr} \rho_{V}^{n} (B)$. By substituting them into the definition (\ref{eq:replicami}) of $I_{n}(A,B)$, and by using the expansion (\ref{eq:exphightw}), we get

\be
I_{n}(A,B) =\f{(4l_{A}l_{B})^{2\Delta}}{4} \sum_{(p_{1},j)=1, p_{1} \neq j}^{n} \sum_{(p_{2},k)=1, p_{2} \neq k}^{n}  \tilde{C}_{(p_{1},j)} \tilde{C}_{(p_{2},k)}\la V^{\otimes n} B_{r_{A}}(p_{1}, j) (\infty)B_{r_{B}} (p_{2}, k) V^{\otimes n}(0) \ra.
\ee

Note that this expression of $I_{n}(A,B)$ is essentially same as that of the 2d counterpart (\ref{eq:mutn}). 
This in particular means that (\ref{eq:inex}) still holds in the higher dimensional case,  once we do the replacement $C(p_{1} -j) \rightarrow \tilde{C}_{(p_{1},j)}$.  In \cite{Agon:2015ftl} it was shown that 
\be
\sum_{k=1, k \neq p_{1}}^{n}  \tilde{C}_{(p_{1},j)}^2  \rightarrow  (n-1) \f{\Gamma(3/2)\Gamma(2\Delta+1)}{2^{4\Delta} \Gamma(2\Delta+3/2)}, \quad n \rightarrow 1,
\ee 

therefore  the result  (\ref{eq:finalresult}) remains to be true in higher dimensions. 
%

\section{Relative entropy of two disjoint intervals}\label{section: hdrel}

In this section we discuss the relative entropy $S_{A\cup B}\;(\rho_{V}|| \rho_{W})$ of two disjoint subsystems  $A \cup B$ in the small subsystem size limit. We take  $\rho_{V}, \rho_{W}$  to be the reduced density matrices of two excited states $|V \rangle, |W \rangle$ on $A \cup B$. We calculate the relative entropy by finding the form of the modular Hamiltonian  $K^{W}_{A \cup B}$ from  entanglement first law, $\delta S= \la  K^{W}_{A \cup B} \; \delta \rho \ra$. Similar trick was used in \cite{relhigerpep} to derive the first asymmetric part of the relative  entropy $S_{A}\;(\rho_{V}|| \rho_{W})$ of a connected region A.

\subsection{ Summary of entanglement entropy}
Here we summarize the result of the entanglement entropy $S_{A \cup B}$ of an exited state $| W \ra$ on  $A \cup B$ in the small interval limit. By using  mutual information $I(A,B)$ we can write this,
\be
S_{A \cup B} =S_{A} +S_{B}- I(A,B). \label{eq:entdis}
\ee

$S_{A}$ is the entanglement entropy of the region A, and  in the small subsystem size limit $l_{A} \rightarrow 0$,   it  is given by (see for example \cite{relhigerpep}), 

\be
S_{A}(\rho_{W}) = \la W| K_{A}^{0}|W \ra -c_{A}\; l_{A}^{2\Delta} \langle W| \mathcal{O}_{A}|W \rangle^2+\cdots  \qquad c_{A} \equiv \f{\Gamma(\f{3}{2}) \Gamma( \Delta+1)}{2 \Gamma(\Delta+\f{3}{2})}. \label{eq:entex}
\ee
where $\cdots$ denotes the terms of $O(l_{A}^{3\Delta})$. $K^{0}_{A}$ is the modular Hamiltonian of the vacuum,  $\mathcal{O}$ is the lightest non vacuum operator, and $\Delta$ is the conformal dimension of the operator  $\mathcal{O}$. 

There is a similar expression for $S_{B}(\rho_{W})$. When the original state is not pure but a mixed state 
\be 
\sigma =\sum_{w} p_{w} |w \ra \la w|,
\ee

then the entanglement entropy of $\sigma$ in this limit is given by 
\be
S_{A}(\sigma) =  {\rm tr} \left[ K^{0}_{A} \sigma \right] -c_{A}\; l_{A}^{2\Delta} \; {\rm tr} \;[ \sigma\mathcal{O}_{A}]^2 +\cdots \label{ent:entmod}.
\ee

The last term in (\ref{eq:entdis}) is the mutual information, in the small subsystem size limit $l_{A}, l_{B} \rightarrow 0$, it  is  given by(\ref{eq:main})
which we reproduce here, 

\begin{align}
I(A,B) =( l_{A})^{2\Delta} ( l_{B})^{2\Delta} \frac{\Gamma(\f{3}{2})\Gamma(2\Delta+1)}{2\Gamma(2\Delta+\f{3}{2})}  \Biggl[ \langle W|  \mathcal{O}_{A} \mathcal{O}_{B}|W \rangle- \langle W| \mathcal{O}_{A} |W\rangle \langle W|\mathcal{O}_{B} |W \rangle \Biggr]^2 + \cdots. \label{ex:mutiex}
\end{align}

By defining 
\begin{align}
c_{AB} &=( l_{A})^{2\Delta} ( l_{B})^{2\Delta} \frac{\Gamma(\f{3}{2})\Gamma(2\Delta+1)}{2\Gamma(2\Delta+\f{3}{2})} \nonumber \\[+10pt]
M^{W}_{AB}&=\langle W | \mathcal{O}_{A}\mathcal{O}_{B}|W \rangle - \langle W|\mathcal{O}_{A} |W \rangle \langle W| \mathcal{O}_{B}|W \rangle,
\end{align}

we can write it as
\be
I^{W}(A,B) =c_{AB}(l_{A},l_{B})( M^{W}_{AB})^2 +\cdots .
\ee

\subsection{ Modular Hamiltonian}

If we slightly deform the reduced density matrix  of $| W \ra$,  $ \rho_{W} \rightarrow \rho_{W}+ \delta \rho$,
then the entanglement entropy changes as
\be
\delta S_{AB} = \la K^{W}_{A \cup B} \; \delta \rho \ra, \quad K^{W}_{A \cup B} = -\log \rho_{W},
\ee
$K^{W}_{A\cup B}$ is the modular Hamiltonian of the state $|W \ra$. 
It is  convenient to divide  it into three parts
\be
K^{W}_{A\cup B} = K^{W}_{A}+K^{W}_{B}+K^{W}_{AB}, \label{eq:modh}
\ee
and we define  each part of the right hand side of (\ref{eq:modh}) by 
\begin{align}
\delta S_{A} = \la K_{A} \delta \rho \ra \quad  \delta S_{B} = \la K_{A} \delta \rho \ra, \label{eq:frista}\\[+10 pt]
-\delta I^{W}(A,B) =\langle K^{W}_{AB}\; \delta \rho  \rangle.
\end{align}

By taking the variation of (\ref{ent:entmod}) with respect to the deformation of $\rho_{W}$ and combine it with (\ref{eq:frista}), we get, 
\be
\delta S_{A} =\la K^{0}_{A} \delta \rho \ra -2c_{A}  l_{A}^{2\Delta} \langle  \mathcal{O}_{A} \rangle_{W} \langle \mathcal{O}_{A} \delta \rho \rangle= \langle K^{W}_{A} \delta \rho \rangle ,  \quad \la \mathcal{O}_{A} \ra_{W} \equiv \la W |\mathcal{O}_{A} |W \ra.
\ee

%


Since the above equation holds for any deformation of the reduced density matrix $\delta \rho$, we can read off the expression of the modular Hamiltonian $K^{W}_{A}$\footnote{ We thank S.Leichenauer for discussion on this.} ,
\be
K^{W}_{A} =K_{A}^{0}-2c_{A} l_{A}^{2\Delta}\langle  \mathcal{O}_{A} \rangle_{W} \mathcal{O}_{A} +\cdots
\ee 
in the small interval limit, $l_{A} \rightarrow 0$.

By using a similar argument for (\ref{ex:mutiex}) , we obtain
\be
K^{W}_{AB} =-2M^{W}_{AB} c_{AB} \Biggl[ \mathcal{O}_{A}\mathcal{O}_{B}-\langle \mathcal{O}_{B} \rangle_{W} \mathcal{O}_{A}  -\langle \mathcal{O}_{A} \rangle_{W} \mathcal{O}_{B} \Biggr] + \cdots. \label{eq:moddis}
\ee


\subsection{Relative entropy of two disjoint interval}

By putting everything together, the relative entropy $S_{A\cup B}(\rho_{V}|| \rho_{W})$ is now given by 

\begin{align}
S_{A\cup B}(\rho_{V}|| \rho_{W}) &= \Delta \langle K^{W}_{A\cup B}  \rangle-\Delta S_{A\cup B} \nonumber  \\[+15pt]
&=\Bigl( \Delta \langle K^{W}_{A} \rangle -\Delta S_{A} \Bigr) +\Bigl( \Delta \langle K^{W}_{B} \rangle -\Delta S_{B}\Bigr) +\Bigl( \langle K^{W}_{AB} \rangle +\Delta I(A,B) \Bigr) \nonumber \\[+20pt]
&= S_{A}(\rho_{V}|| \rho_{W}) + S_{B}(\rho_{V}|| \rho_{W})+c_{AB} \Bigl[ (M^{V}_{AB}-M^{W}_{AB})^2-2 M^{W}_{AB} 
\Bigl( \langle \mathcal{O} \rangle_{V}-\langle \mathcal{O} \rangle_{W} \Bigr)^2 \Bigr] .\label{eq:result}
\end{align}

$S_{A}(\rho_{V}|| \rho_{W}) $ denotes the relative entropy of the region A \cite{Sarosi:2016oks,relhigerpep}, which is given by 
\be
S_{A}(\rho_{V}|| \rho_{W})=c_{A} l_{A}^{2\Delta} \left( \la V | \mathcal{O}_{A} | V \ra - \la W | \mathcal{O}_{A} | W \ra \right)^{2} + \cdots,
\ee
in the small subsystem size limit. We have a similar expression for $S_{B}(\rho_{V}|| \rho_{W})$.

The last two  terms of  (\ref{eq:result})  involve both region A and B.  Note that the last term 
is asymmetric under the exchange of two states $ | V \rangle \leftrightarrow |W \rangle $.


%
%
%
%

\section{Conclusions} 

In the first part of this paper, we derived a general  formula for the mutual information $I(A,B)$ of an arbitrary excited state in the small subsystem size limit.  In the context of holography, this result should agree with 
the mutual information $I( \hat{A}, \hat{B}) $ of  the bulk quantum field theory of the regions $\hat{A}, \hat{B}$ which are enclosed by the  corresponding boundary subsystems and the  bulk RT surfaces \cite{Ryu:2006bv,Ryu:2006ef}, according to the FLM conjecture \cite{Faulkner:2013ana}. It would be interesting to check this statement concretely. When the CFT state is vacuum, this was explicitly confirmed in \cite{Agon:2015ftl} by calculating the mutual information on the dual geometry, ie, anti de Sitter space.  To reproduce our result (\ref{eq:main}) by a bulk calculation,  we need to generalize their work to the asymptotically AdS space with the back reaction of the scalar condensate dual to $\la \mathcal{O} \ra$.   

\vspace{0.1cm}

In the second part, we derived a formula for the relative entropy $S_{A\cup B} (\rho_{V} || \rho_{W})$ of disjoint subsystems in the small subsystem size limit. We did this by reading off the form of the modular Hamiltonian 
from the entanglement first law.
It will be an interesting future work to study the relative entropy numerically.

\section*{Acknowledgments}

We thank Joan Simon and Gabor Sarosi  for initial collaboration, David Williams and  Hiroshi Ooguri for discussions.  Part of this work was done during the YITP long term workshop "Quantum Information in String Theory and Many-body Systems".  We would like to thank 
the institute for its hospitality. 
The work of  T. U. was supported in part by the National Science Foundation under Grant
No. NSF PHY-25915.

\appendix

\section{Some details of the calculation of  $I_{n} (A,B)$}
In this appendix  we explain the detail of the calculation of  $I_{n} (A,B)$ defined by
\be
I_{n}(A:B)=\f{1}{4} (l_{12}l_{34})^{2\Delta} \sum_{(j,p_{1}) =1, j \neq p_{1} }^{n} \sum_{(k,p_{2}) =1, k \neq p_{2} }^{n} C^{B(j , p_{1})}_{\sigma_{n}\tilde{\sigma}_{-n}} C^{B(k, p_{2})}_{\sigma_{n}\tilde{\sigma}_{-n}} 
\; \langle V_{n}(\infty) B_{z_{2}}(p_{1}, j) B_{z_{4}} (p_{2}, k) V_{n}(0) \rangle .
\ee

To do this we  first  compute the sum,
\be
I_{p_{1} p_{2}}=\sum_{j=1, k \neq p_{1}}^{n}\sum_{k=1, k \neq p_{2}}^{n} \; \langle V_{n}(\infty) B_{z_{2}}(p_{1}, j) B_{z_{4}} (p_{2}, k) V_{n}(0) \rangle \; C_{\sigma_{n}\tilde{\sigma}_{-n}}^{B(j,p_{1})} C_{\sigma_{n}\tilde{\sigma}_{-n}}^{B(k,p_{2})} \equiv \sum_{j=1, k \neq p_{1}}^{n}\sum_{k=1, k \neq p_{2}}^{n} I^{k,j}_{p_{1} p_{2}},
\ee

 for fixed $p_{1},p_{2}$, by classifying the possible forms of the summand $I^{j,k}_{p_{1}, p_{2}}$, then performing the  sum with respect to  $p_{1},p_{2}$.

\subsubsection{ $ I_{p_{1},p_{2}}:p_{1} \neq p_{2}$ case}

This case is further classified into four possibilities, $\{(j=p_{2}), (k=p_{1})\}$, 
$\{(j \neq p_{2}), (k=p_{1})\}$, $\{(j=p_{2}), (k\neq p_{1})\}$, $\{(j \neq p_{2}), (k\neq p_{1})\}$. 

\vspace{0.4cm}

{\bf \underline{$\{(j=p_{2}), (k=p_{1})\}$ }}

\vspace{0.4cm}

When   $\{(j=p_{2}), (k=p_{1})\}$ the summand in (\ref{eq: fourpt}) is given by
\be
I^{p_{2},p_{1}}_{p_{1} p_{2}} =C(p_{1}-p_{2})^2  \langle  \mathcal{O}_{\alpha} \mathcal{O}_{\beta} \rangle^{2},\label{eq:r1}
\ee
here we introduced  the simplified notations,
\be
C(p_{1}-p_{2}) \equiv  C_{\sigma_{n}\tilde{\sigma}_{-n}}^{T_{(p_{1}-p_{2})}} =C^{B(p_{1},p_{2})}_{\sigma_{n} \tilde{\sigma}_{-n}},\quad   \langle \mathcal{O}_{\alpha} \mathcal{O}_{\beta} \rangle \equiv \langle V(\infty)  \mathcal{O} (z_{2})\mathcal{O} (z_{4}) V(0) \rangle.
\ee

\vspace{0.4cm}

{\bf \underline{$\{(j=p_{2}), (k \neq p_{1})\}$ }}

\vspace{0.4 cm}

When  $\{(j=p_{2}), (k \neq p_{1})\}$, The summand is given by 
\be
I^{p_{2},k}_{p_{1},p_{2}}= C(p_{2}-p_{1}) C(k-p_{2}) \langle  \mathcal{O}_{\alpha} \mathcal{O}_{\beta} \rangle \langle  \mathcal{O}_{\alpha} \rangle \langle \mathcal{O}_{\beta} \rangle,
\ee
with 
\be
 \langle  \mathcal{O}_{\alpha} \rangle = \langle V(\infty)  \mathcal{O}(z_{2}) V(0) \rangle, \quad  \langle  \mathcal{O}_{\beta} \rangle = \langle V(\infty)  \mathcal{O}(z_{4}) V(0) \rangle.
\ee
The sum with respect to $k$ is given by,
\begin{align}
\sum_{k=1,k \neq p_{1},p_{2}}^{n} I^{p_{2},k}_{p_{1} p_{2}} &= \sum_{k=1,k \neq p_{1},p_{2}}^{n} C(p_{1}-p_{2}) C(k-p_{2}) \langle  \mathcal{O}_{\alpha} \mathcal{O}_{\beta} \rangle \langle  \mathcal{O}_{\alpha} \rangle \langle \mathcal{O}_{\beta} \rangle  \\
&= C(p_{1}-p_{2})\left[  \sum_{k=1,k \neq p_{2}}^{n} C(k-p_{2}) -C(p_{1}-p_{2}) \right]  \langle  \mathcal{O}_{\alpha} \mathcal{O}_{\beta} \rangle \langle  \mathcal{O}_{\alpha} \rangle \langle \mathcal{O}_{\beta} \rangle. \label{eq:r2}
\end{align}

\vspace{0.4cm}

\underline{ $\{(j\neq p_{2}), (k= p_{1})\}$}

\vspace{0.4cm}

Similarly,  when $\{(j\neq p_{2}), (k= p_{1})\}$

\be 
\sum_{j=1, j \neq p_{1}, p_{2}}^{n} I^{j,p_{1}}_{p_{1},p_{2}} =C(p_{1}-p_{2})\left[  \sum_{j=1, j \neq p_{1}}^{n} C(j-p_{1}) -C(p_{1}-p_{2}) \right]  \langle  \mathcal{O}_{\alpha} \mathcal{O}_{\beta} \rangle \langle  \mathcal{O}_{\alpha} \rangle \langle \mathcal{O}_{\beta} \rangle .\label{eq:r3}
\ee

\vspace{0.4cm}

{\bf \underline{$\{(j \neq p_{2}), (k \neq p_{1})\}$ }}

\vspace{0.4 cm}

Let us finally consider the $\{(j \neq p_{2}), (k\neq p_{1})\}$ case. It is useful to further separate this case into  $j=k$, and $j \neq k$. The $j=k$ result is given by 
\be 
\sum_{k=1, k\neq p_{1},p_{2}} I^{k,k}_{p_{1},p_{2}}=\sum_{k=1, k\neq p_{1},p_{2}} C(k-p_{1})C(k-p_{2}) \langle  \mathcal{O}_{\alpha} \mathcal{O}_{\beta} \rangle \langle  \mathcal{O}_{\alpha} \rangle \langle \mathcal{O}_{\beta} \rangle. \label{eq:r4}
\ee

The  $j \neq k$ result is 
\begin{flalign}
\sum_{j=1,j\neq p_{1}}^{n} \sum_{k=1,k \neq p_{1}, j}^{n} I^{j,k}_{p_{1},p_{2}} &=\sum_{j=1,j\neq p_{1}}^{n} \sum_{k=1,k \neq p_{1}, j}^{n} C(j-p_{1})C(k-p_{2}) \langle  \mathcal{O}_{\alpha} \rangle^{2} \langle \mathcal{O}_{\beta} \rangle^{2}& \nonumber \\
&=\left[ \left( \sum_{j=1, j \neq p_{1}}^{n} C(j-p_{1}) \right)\left( \sum_{k=1, k \neq p_{2}}^{n} C(k-p_{2}) \right) -\sum_{k=1, k\neq p_{1},p_{2}}^{n} C(k-p_{1})C(k-p_{2}) \right.&\nonumber \\
&\left.-C(p_{1}-p_{2}) \left( \sum_{j=1, j \neq p_{1}}^{n} C(j-p_{1}) +\sum_{k=1, k\neq p_{2}}^{n} C(k-p_{1})-C(p_{1}-p_{2})\right) \right]
 \langle \mathcal{O}_{\alpha} \rangle^{2} \langle \mathcal{O}_{\beta} \rangle^{2}.& \label{eq:r5}
\end{flalign}

\vspace{0.4cm}

\underline{{\bf Net result for $I_{p_{1} p_{2}}$}}

\vspace{0.4 cm}

By putting  (\ref{eq:r2}), (\ref{eq:r3}), (\ref{eq:r4}), (\ref{eq:r5}) together, we get,
\begin{flalign}
I_{p_{1},p_{2}}&=\sum_{j=1, k \neq p_{1}}^{n}\sum_{k=1, k \neq p_{2}}^{n} I^{j,k}_{p_{1}, p_{2}} =C(p_{1}-p_{2})^2 \Bigl[ \langle  \mathcal{O}_{\alpha} \mathcal{O}_{\beta} \rangle -\langle O_{\alpha}\rangle \langle O_{\beta}\rangle \Bigr]^2 &\nonumber \\
&+\left[ \left( \sum_{k=1, k \neq p_{1},p_{2}}^{n}C(k-p_{1})C(k-p_{2}) \right)+C(p_{1}-p_{2}) \left( \sum_{k=1, k\neq p_{2}}^{n} C(k-p_{2}) + \sum_{j=1, j\neq p_{1}}^{n} C(j-p_{1})\right) \right] &\\
& \times \Bigl[ \langle  \mathcal{O}_{\alpha} \mathcal{O}_{\beta} \rangle -\langle  \mathcal{O}_{\alpha} \rangle \langle \mathcal{O}_{\beta} \rangle \Bigr] \langle  \mathcal{O}_{\alpha} \rangle \langle \mathcal{O}_{\beta} \rangle \nonumber&\nonumber \\[+10pt] 
&\rightarrow  \; C(p_{1}-p_{2})^2 \Bigl[ \langle  \mathcal{O}_{\alpha} \mathcal{O}_{\beta} \rangle -\langle O_{\alpha}\rangle \langle O_{\beta}\rangle \Bigr]^2 & \nonumber \\
&+ \sum_{k=1, k \neq p_{1},p_{2}}^{n} C(k-p_{1})  C(k-p_{2}) \Bigl[ \langle  \mathcal{O}_{\alpha} \mathcal{O}_{\beta} \rangle -\langle  \mathcal{O}_{\alpha} \rangle \langle \mathcal{O}_{\beta} \rangle \Bigr]\langle  \mathcal{O}_{\alpha} \rangle \langle \mathcal{O}_{\beta} \rangle ,\quad n \rightarrow 1.   \label{eq:neq}
\end{flalign}
In the last line we only picked up terms which are proportional to (n-1) after the final analytic continuation. 
Terms which are ignored in the last line are proportional to $(n-1)^{2}$ in $ I_{n}(A:B)$. For example  
\be
\sum_{p_{1}=1, p_{1} \neq p_{2}}^{n} \sum_{k=1, k\neq p_{2}}^{n} C(p_{1}-p_{2}) C(k-p_{2}) = f(n, \Delta/2)^{2} \propto (n-1)^2,  \quad n \rightarrow 1.
\ee
where  $f(n, \Delta)$ is defined in (\ref{eq:anacon}),
%
%
\subsubsection{ $I_{p_{1},p_{2}}: p_{1} = p_{2}$ case}

\vspace{0.3cm}

When $j=k$,
\be
 \sum_{k=1, k\neq p_{1}}^{n} I^{k,k}_{p_{1},p_{1}} = \sum_{k=1, k\neq p_{1}}^{n} C(p_{1}-k)^2  \langle \mathcal{O}_{\alpha} \mathcal{O}_{\beta} \rangle^2.
\ee

 $j \neq k$ result is
\begin{align}
 \sum_{j=1, j\neq p_{1}}^{n}\sum_{k=1, k \neq j, p_{1}}^{n} I^{k,j}_{p_{1},p_{1}} &= \sum_{j=1, j\neq p_{1}}^{n}\sum_{k=1, k \neq j, p_{1}}^{n} C(j-p_{1}) C(k-p_{1}) \langle  \mathcal{O}_{\alpha} \mathcal{O}_{\beta} \rangle \langle  \mathcal{O}_{\alpha} \rangle \langle \mathcal{O}_{\beta} \rangle \\
&=\left[\left( \sum_{k=1, k\neq p_{1}}^{n} C(k-p_{1}) \right)^2 -\sum_{k=1, k\neq p_{1}}^{n} C(k-p_{1})^2 \right] \langle  \mathcal{O}_{\alpha} \mathcal{O}_{\beta} \rangle \langle  \mathcal{O}_{\alpha} \rangle \langle \mathcal{O}_{\beta} \rangle.
\end{align}

Therefore
\begin{align}
I_{p_{1},p_{1}} 
 \rightarrow \sum_{k=1, k\neq p_{1}}^{n} C(p_{1}-k)^2 \Bigl[ \langle  \mathcal{O}_{\alpha} \mathcal{O}_{\beta} \rangle- \langle \mathcal{O}_{\alpha} \rangle \langle \mathcal{O}_{\beta} \rangle \Bigr]  \langle \mathcal{O}_{\alpha}  \mathcal{O}_{\beta} \rangle. \label{eq:equal}
\end{align}
Again we only picked up terms surviving $n \rightarrow 1$ limit.

\subsubsection{The final result}

By combining (\ref{eq:equal}) and  (\ref{eq:neq}), and using the identity 
\be 
\sum_{p_{1}\neq p_{2}} \sum_{k \neq p_{1},p_{2}} C(k-p_{1})C(k-p_{2}) =n\left(\sum_{p_{1}=1, p_{1}\neq k} C(k-p_{1}) \right)^{2}-\sum_{k} \sum_{p_{1} \neq k} C(p_{1}-k)^2,
\ee

we get the  expression of  $I_{n}(A:B)$,
\begin{align}
 I_{n}(A:B)&= \f{1}{4} (l_{12})^{\Delta} (l_{34})^{\Delta} \left(\sum _{p_{1} \neq p_{2}} I_{p_{1}, p_{2}} + \sum_{p_{1}} I_{p_{1}, p_{1}} \right) \\[+10pt]
&= \f{1}{2}  (l_{12})^{\Delta} (l_{34})^{\Delta} \left[  \langle  \mathcal{O}_{\alpha} \mathcal{O}_{\beta} \rangle- \langle \mathcal{O}_{\alpha} \rangle \langle \mathcal{O}_{\beta} \rangle \right]^2  \sum_{k=1,k\neq p_{1}}^{n} C(p_{1}-k)^2.
\end{align}
%

{100}

\end{document}